\documentclass[10pt,letterpaper]{article}
\usepackage[utf8]{inputenc}
\usepackage{amsmath}
\usepackage{amsfonts}
\usepackage{amssymb}
\usepackage{graphicx}
\begin{document}
\title{Dark Energy and Dark Matter in $f(R)$ Gravity and an Alternative Chameleon Mechanism}
\author{Adam Kelleher}
\maketitle
\abstract{Metric $f(R)$ gravity theories behave like GR with cosmological constant when $f''(R)$ = 0, and like scalar-tensor theories elsewhere.  I investigate the transition from $f''(R) \neq 0$ to $f''(R) = 0$, and show that this theory may offer a way to explain Dark Energy and/or Dark Matter without having to make use of the chameleon mechanism to agree with solar system tests.}

\section{Introduction}

There is a large literature attempting to use $f(R)$ gravity to explain dark energy and dark matter, as reviewed in \cite{DeFelice2010}\cite{Sotiriou}.  The attempts to explain dark energy made use of the chameleon mechanism \cite{KhouryWeltman} in order to agree with solar system experiments (e.g. \cite{Lombriser2013}\cite{Tsujikawa2008}\cite{Capozziello2008}), but this mechanism does not appear to be satisfactory, in general.  The chameleon mechanism applied to metric $f(R)$ gravity results in apparent fifth forces which are large, and should be measurable in principle \cite{HuiNicolisStubbs}.  The strength of the chameleon mechanism was, in part, that it came out simply from the theory, and applied to a large class of $f(R)$.  If we have to add a lot of complexity to a theory to make it work with experiments, we should grow very skeptical of the theory.  I will show that there is a rather simple replacement for the chameleon mechanism, in the sense that it suppresses solar system effects in $f(R)$ gravity, and again applies to a broad class of forms for $f(R)$.

This theory has also been applied in an attempt to describe dark matter (e.g.\cite{Boehmer2007}\cite{Lubini2011}\cite{Capoziello2012}). While it is easy to choose a form for $f(R)$ that gives its effective degree of freedom an equation of state like regular matter, it is difficult in practice to choose a form for $f(R)$ that agrees with experimental results \cite{Sotiriou}.  In particular, empirically fitting parameterized models for $f(R)$ to galaxy lensing data produces a fit that depends on the masses of the galaxies \cite{Sotiriou}\cite{Mendoza2006}.  I believe it might be worth looking at this work again, given the results presented in this paper.  
 
The key to the mechanism I am proposing is that $f(R)$ gravity can behave very differently in regions of different scalar curvature.  In particular, when $f(R) \sim R$, the theory behaves just like general relativity.  When $f(R)$ is different from $R$, the theory effectively picks up an extra scalar degree of freedom.  I will show this in detail in the following section.  The main point is that we can now construct a form for $f(R)$ that agrees with solar system experiments due to the fact that $f(R) \sim R$ in regions of background curvature greater than or equal to that in the solar system, and different from $R$ outside the solar system.  Doing this will result in novel gravity wave reflection effects.

In the next section, I will outline how the scalar degree of freedom can be added and removed.  Then, I will go on to construct a toy $f(R)$ to illustrate that it is possible.  I'll then describe some new phenomena implied by this mechanism, and conclude with a summary of these results and their implications.

\section{Scalar Degree of Freedom}

There is an equivalence between scalar-tensor theory that applies whenever $f''(R) \neq 0$.  When this condition is violated, the equivalence breaks down.  The reason for this becomes clear when we solve the equation $f''(R) = 0$.  The result, of course, is just $f(R) = R + const.$.  Thus, the theory reduces to GR with a cosmological constant, and we have no extra scalar degree of freedom.

We can ask what happens when we do not quite reach $f''(R) = 0$, but instead just come very close to it.  Examining the dynamic equation for scalar perturbations, we find that the mass increases like $1/f''(R)$.  Thus, the mass increases, and scalar mode production is strongly suppressed, and the range of the scalar force is strongly suppressed.

We can consider, then, constructing a form for $f(R)$ where for some range of $R$ values, we are very close to $f''(R) = 0$.  This implies that in any spatial region where the curvature falls into this range of values, the modifications to GR are strongly suppressed.  

To make the point more concrete, I will take a particular example for $f(R)$ with a well-defined point $R_0$ where the theory changes from GR to modified gravity.  I will then introduce a physical scenario where we look at the theory as we move in one direction through a gas with slowly decreasing density.  As this density decreases, so does the scalar curvature in the region it occupies.  Thus, we can see the behavior caused by changing $f(R)$ by moving through space.

\section{Discontinuous Case}
\begin{figure}\label{fig:5-1_2}
\begin{center}
\includegraphics[width=0.5\textwidth]{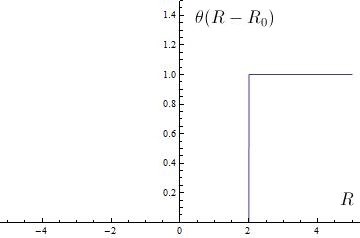}
\end{center}
\caption{A displaced $\theta$ function, $\theta(R - R_0)$.}  
\end{figure}

\begin{figure}\label{fig:5-1_3}
\begin{center}
\includegraphics[width=0.5\textwidth]{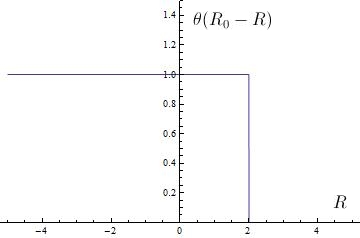}
\end{center}
\caption{A reflected, displaced $\theta$ function, $\theta(R_0 - R)$.}  
\end{figure}

$\indent$First, recall the Heaviside $\theta$ function, $\theta(x)$,
\begin{eqnarray}
\theta(x) = 
1 & x > 0 \nonumber  \\
0 & x \leq 0 \nonumber \\
.
\end{eqnarray}
I'll review some basic manipulations, since we'll go into some depth with them.  The graph of a displaced $\theta$ function, $\theta( R - R_0)$ is shown in figure \ref{fig:5-1_2}.  If we reverse the sign, it reflects the function across $R = R_0$, as shown in figure \ref{fig:5-1_3}.
\begin{figure}\label{fig:5-1_1.jpg}
\begin{center}
\includegraphics[width=0.5\textwidth]{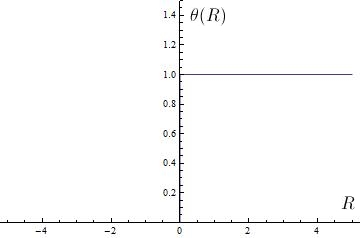}
\end{center}
\caption{A $\theta$ function, $\theta(R)$.}  
\end{figure}

\begin{figure}\label{fig:5-1_a}
\begin{center}
\includegraphics[width=0.5\textwidth]{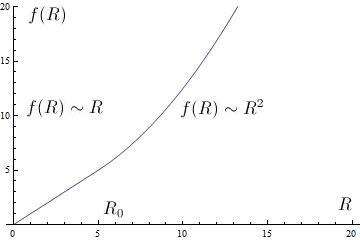}
\end{center}
\caption[An $f(R)$ with a critical point]{The function $f(R)$, where below $R_0$ the theory is like GR.  Above $R_0$, there is another scalar degree of freedom.}  
\end{figure}

For $f(R)$, we will choose the function
\begin{equation}
f(R) = R + \theta(R - R_0) (\mathcal{ F } (R) - R),
\end{equation}
whose graph is depicted in figure \ref{fig:5-1_a}.

\begin{figure}\label{fig:5-1_b}
\begin{center}
\includegraphics[width=0.5\textwidth]{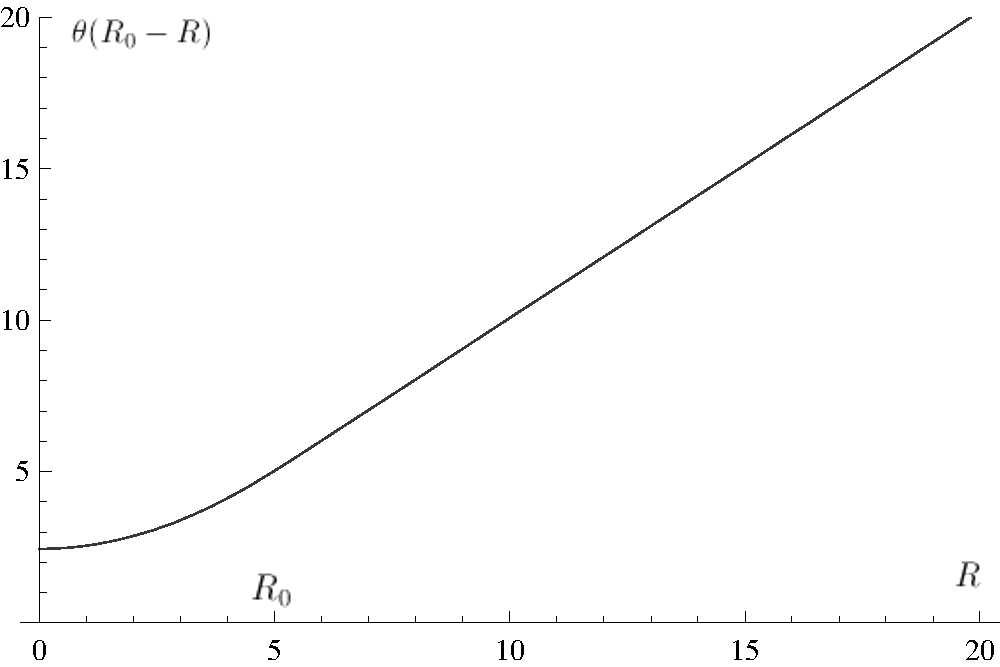}
\end{center}
\caption{The function $f(R)$, where above $R_0$ the theory is like GR.  Below $R_0$, there is another scalar degree of freedom.}  
\end{figure}

Examining the above function, we see $R < R_0$ implies $f(R) = R$.  Also, $R \geq R_o$ implies $f(R) = \mathcal{F}(R)$.  We have imposed a cutoff at $R_o$ where our theory abruptly changes from GR to modified gravity, and $f''(R) = 0$ for $R > R_o$, but is left general otherwise.  Before getting into details about continuity and differentiability, lets say exactly what we mean by ``acts like GR (or modified gravity)" for certain values of R.  We can reverse this and put the modified gravity effects at higher curvatures, while keeping GR at lower curvatures, by flipping the sign of the $\theta$ function.  The resulting $f(R)$ is shown in figure \ref{fig:5-1_b}

Let's examine the field equations for metric $f(R)$ gravity.  We will write them again here for convenience,
\begin{eqnarray}
f'(R) R_{\mu \nu} - \frac{1}{2}f(R) g_{\mu \nu} - \left[ \nabla_{\mu}\nabla_{\nu} - g_{\mu \nu} \square \right] f'(R) & = & \kappa T_{\mu \nu}.
\end{eqnarray}
When $f(R) = R + const.$, these field equations reduce to GR with cosmological constant.  This is the situation implied when $f''(R) = 0$.  When this constraint is met over some range of $R$, then the field equations are effectively the GR field equations in that range of curvatures.  

This point becomes more clear when we examine the trace equation,
\begin{equation}
f'(R) R - 2 f(R) + 3 \square f'(R) = \kappa T,
\end{equation}.
and recall that when we make the equivalence with scalar-tensor theory, this is the wave equation for the field $\phi = f'(R)$.  If we imagine that these scalar waves are a perturbation to the background curvature, $R_b$ (i.e. that $R = R_b + \delta R$), then we can write $\square f'(R) \simeq f''(R_b) \square \delta R$, and it becomes clear that the constraint $f''(R) = 0$ being true in some range of $R$ implies that there are no dynamic scalar wave perturbations to that range of background curvatures.

\section{Continuous Case}
$\indent$Now, we can confront the issue of continuity and differentiability.  Because of the Heaviside function used to switch between theories, we can get ugly artifacts.  In particular, if $f(R)$ isn't differentiable, we can't include $f'(R)$ in our field equations.  We must have well-defined field equations, so we need $f(R)$ to be differentiable.  I will now show that, as the reader might expect, this problem is not fundamental to the choice $f''(R) = 0$, but is simply an artifact of our use of the Heaviside function.

\begin{figure}\label{fig:5-2_1}
\begin{center}
\includegraphics[width=0.5\textwidth]{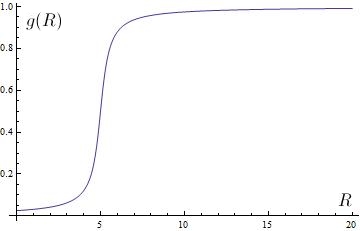}
\end{center}
\caption[An approximate $\theta$ function]{The function $g(R)$, which is a smoothed displaced, reflected $\theta$ function, whose transition is at $R_0$.}  
\end{figure}

Let us use a somewhat more complicated function in place of the Heaviside function.  We replace $\theta(R_0 - R)$ with $g(R)$, defined by
\begin{equation}
g(R) = \frac{1}{2}\left[ 1 - \frac{2}{\pi} \,\mathrm{atan}\,\left(\frac{R/R_0 - 1}{\epsilon}\right) \right]
\end{equation}, which is graphed in figure \ref{fig:5-2_1}.

Notice that as $x = (R/R_o - 1)/ \epsilon \rightarrow \infty$, $g(R)$ approaches 0.  As $x \rightarrow -\infty$, $g(R)$ approaches 1.  As $\epsilon \rightarrow 0$, $g(R)$ simply becomes the Heaviside function, $\theta(R_o - R)$.  We can use this to get rid of the discontinuity in our choice of $f(R)$, while still suppressing modifications to gravity in certain ranges of $R$.  Let's examine the asymptotics for $g(R)$.

First, we note the asymptotics for $\mathrm{atan}(x)$.  For $x \rightarrow -\infty$, we find $\mathrm{atan}(x) = -\frac{\pi}{2} + \frac{1}{x} - \frac{1}{3 x^3} + \, \cdots$.  Similarly, as $x \rightarrow \infty$ we find $\mathrm{atan}(x) = \frac{\pi}{2} - \frac{1}{x} + \frac{1}{3 x^3} - \,\cdots$.  Plugging these in for $g(R)$ for small and large $x = (R/R_o - 1)/\epsilon$, we find for large $x$, 
\begin{equation}
g(x \gg 1) = \frac{\epsilon}{\pi(R/R_o - 1)} - \frac{\epsilon^3}{3 \pi (R/R_o - 1)^3} + \, \cdots 
\end{equation}
and for $x \ll 1$,
\begin{equation}
g(x \ll -1) = 1 - \frac{\epsilon}{\pi(R/R_o - 1)} + \frac{\epsilon^3}{3 \pi (R/R_o - 1)^3} - \, \cdots
\end{equation}.

Thus, as we increase the sharpness of our cutoff (smaller $\epsilon$), or get farther from it, we approach the Heaviside case.  

\begin{figure}\label{fig:5-2_2}
\begin{center}
\includegraphics[width=0.5\textwidth]{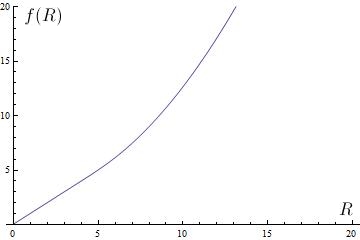}
\end{center}
\caption{The differentiable function, $f(R)$.}  
\end{figure}
Going farther, we can see where the corrections to the field equations come in.  We can define a new, continuous $f(R)$ with this new $g(R)$, as shown in figure \ref{fig:5-2_2}.  First, we calculate the large and small $x$ asymptotics of $f(R)$,

\begin{equation}
f(x \ll -1) = \mathcal{F}(R) + (\mathcal{F}(R) - R)\left( - \frac{\epsilon}{\pi(R/R_o - 1)} + \frac{\epsilon^3}{3 \pi (R/R_o - 1)^3} - \, \cdots \right)
\end{equation}
and 
\begin{equation}
f(x \gg 1) = R + (\mathcal{F}(R) - R)\left( \frac{\epsilon}{\pi(R/R_o - 1)} - \frac{\epsilon^3}{3 \pi (R/R_o - 1)^3} + \, \cdots \right)
\end{equation}.

These are both corrections to the Heaviside version at the order $1/x$ for $\lvert x \rvert \gg 1$.  Continuing, we calculate asymptotics for $f'(R)$ as 
\begin{equation}
f'(x \ll -1) = \mathcal{F}'(R) + (\mathcal{F}'(R) - 1) \left( - \frac{\epsilon}{\pi(R/R_o - 1)} + \frac{\epsilon^3}{3 \pi (R/R_o - 1)^3} - \, \cdots \right)
\end{equation}
and 
\begin{equation}
f'(x \gg 1) = 1 - \frac{\epsilon}{\pi R_o}\left( \frac{1}{1 + x^2 } \right) \left( \mathcal{F}(R) - R \right) + \left( \frac{1}{\pi x} - \frac{1}{3 \pi x} \right)\left( \mathcal{F}'(R) - 1 \right) + \, \cdots \label{fpxggo}
\end{equation}

We can put all of this into our expression for the field equations, and see that corrections come in at order $1 / x$.  For example, we can examine the trace equation.  This will show us at what order the dynamics for $\delta R$ start to come in.  Examining (\ref{fpxggo}), we see that the wave operator term is suppressed by at least $1/x$.  By increasing the sharpness of our cutoff, or by moving sufficiently far from it, we can suppress this term to arbitrarily small values.  This effectively makes the mass of these perturbations too large to ever hope to excite one.

\section{New Phenomena Near $f''(R) = 0$}
\begin{figure}\label{fig:5-3_1}
\begin{center}
\includegraphics[width=0.5\textwidth]{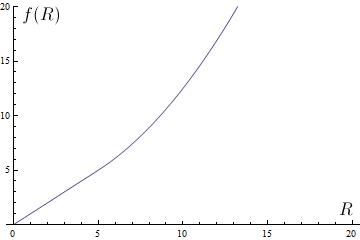}
\end{center}
\caption[A differentiable $f(R)$ with critical point effects]{The differentiable function, $f(R)$, used to illustrate scalar wave effects.  As the scalar wave moves from a region of high curvature to a region of lower curvature, it can be reflected at the boundary where $R \rightarrow R_0$.}  
\end{figure}

$\indent$Something very interesting can happen when the scalar field theory equivalent to $f(R)$ gravity is examined near a region where $f''(R) \rightarrow 0$.  I'm not aware of this having been written about before.  In this region, it has been known that the mass of the scalar can be large, since $m_{\phi} \propto (f''(R))^{-1}$.  This has been examined before in the context of the chameleon mechanism, where scalar effects are suppressed near massive bodies.  Consider a case where the scalar curvature is only slightly perturbed from its GR value, $R = - \kappa T + R_1$, and for low energies $T$ is approximately proportional to the energy density, $T = -\rho$.  Now, consider a form of $f(R)$ where $f''(R) \geq 0$ for $R \geq R_0$, but $f''(R) < \delta$ for $R < R_0$, for some small parameter $\delta$, as depicted in figure \ref{fig:5-3_1}.  If we examine some energy distribution $\rho$ where $R = \kappa \rho + R_1 \geq R_0$ in some sphere with radius $r < r_0$, and  $R = \kappa \rho + R_1 < R_0$ for $r \geq r_0$, we see something very interesting happening.  Inside the sphere, the scalar mass is some value away from $0$.  Outside, however, the scalar mass gets large very fast.  In this region, the scalar is not able to propagate, and we get reflection at the boundary $r = r_0$.  We can see this clearly if we examine the perturbation equations for the  scalar degree of freedom (\cite{Faraoni2006}),

\begin{eqnarray}\label{scalarperteqn}
& & \ddot{R_1} - \mathbf{\nabla}^2 R_1 - \frac{2 \kappa \varphi'''}{\varphi''}\dot{T}\dot{R_1} + \frac{2 \kappa \varphi'''}{\varphi''} \mathbf{\nabla}T \cdot \mathbf{\nabla}R_1 \\
& & + \frac{1}{3 \varphi''}\left( \frac{1}{\epsilon} - \varphi' \right) R_1 = \kappa \ddot{T} - \kappa \mathbf{\nabla}^2 T - \frac{\kappa T \varphi' + 2 \varphi}{3 \varphi''} \nonumber.
\end{eqnarray}

Under the approximations

\begin{eqnarray}\label{perturbapprox}
f'(R) & = & 1 + \epsilon \varphi' \nonumber \\
\square f'(R) & = & \epsilon \varphi'' \square R  \nonumber \\
g_{\mu \nu} &  =  & \eta_{\mu \nu} + h_{\mu \nu} \nonumber \\
R & = & -\kappa T + R_1, \nonumber
\end{eqnarray}
we find that the equations of motion for the scalar degree of freedom simplify to 
\begin{eqnarray}
& & \ddot{R_1} - \mathbf{\nabla}^2 R_1 - \frac{2 \kappa \varphi'''}{\varphi''}\dot{T}\dot{R_1} + \frac{2 \kappa \varphi'''}{\varphi''} \mathbf{\nabla}T \cdot \mathbf{\nabla}R_1 \nonumber \\
& & + \frac{1}{3 \varphi''}\left( \frac{1}{\epsilon} - \varphi' \right) R_1 = \kappa \ddot{T} - \kappa \mathbf{\nabla}^2 T - \frac{\kappa T \varphi' + 2 \varphi}{3 \varphi''}, \nonumber
\end{eqnarray}
where $\varphi$ and its derivatives are evaluated at $R = -\kappa T$.

Since we're really only interested in a region close to the boundary, we can choose somewhere where $T$ varies only very slightly.  We will still get very abrupt reflection effects.  We can also choose to work with a static background.  This makes derivatives of $T$ all vanish.  To simplify to the essence of the effect, let's restrict to one dimension.  This reduces our equations of motion to a one-dimensional wave equation,

\begin{equation}\label{scalareffeqn}
\frac{\partial^2 R_1}{\partial t^2} - \frac{\partial^2 R_1}{\partial x^2} + \frac{1}{3 \varphi ''}\left( \frac{1}{\epsilon} - \varphi' \right)R_1 = - \frac{\kappa T \varphi' + 2 \varphi}{3 \varphi''}
\end{equation}

Here we run into a (surmountable) issue.  Naively looking for small wave effects, and small changes from GR values puts us in an awkward position when we expect large values.  This is the case when we're asking what the background scalar field value is in a region of uniform density.  In such a region, the scalar curvature has a constant added to it due to a contribution from the scalar field.  This makes sense, as we know that the scalar field will tend to a constant background value in the presence of matter.  Realizing this, we can modify the approach taken in \cite{Faraoni2006} to allow for this extra contribution from the scalar field.  We can generalize the equations (\ref{perturbapprox}) to
\begin{eqnarray}
f'(R) & = & 1 + \epsilon \varphi ' \\
f(R) & = & -\kappa T + R_s + R_1 + \epsilon \varphi \\
R & = & -\kappa T + R_s + R_1
\end{eqnarray}

Here, $R_s$ represents the static contribution to the Ricci curvature due to the scalar field, and $R_1$ will end up representing wave perturbations.  To find our value for $R_s$, we have to examine the equation of motion in $f(R)$ gravity representing the scalar degree of freedom.  We will look at it in the geometric form,
\begin{equation}
3 \square f'(R) + f'(R) R - 2 f(R) = \kappa T.
\end{equation}

Plugging in our new perturbation scheme,  we get 
\begin{equation}
( 1 + \epsilon \varphi')(-\kappa T + R_s + R_1)  - 2( -\kappa T + R_s + R_1 + \epsilon \varphi) = \kappa T
\end{equation}
which simplifies to
\begin{equation}
\varphi' \kappa T + 2 \varphi  =  \left( \varphi' - \frac{1}{\epsilon} \right) R_s.
\end{equation}

We can divide through by $-1/3 \varphi''$ to put this in the much more useful form,
\begin{equation}\label{R_sterm}
-\frac{\kappa T \varphi' + 2 \varphi}{3 \varphi''} = \left( \frac{1}{\epsilon} - f'(R) \right) \frac{1}{3 \varphi''} R_s
\end{equation}

Now it is probably apparent that this term will end up eliminating the source term for our wave equation.  If we apply the new perturbation scheme to the original wave equation, we get
\begin{equation}\label{newwaveeqn0}
\frac{\partial^2 R_1}{\partial t^2} - \frac{\partial^2 R_1}{\partial x^2} + \frac{1}{3 \varphi ''}\left( \frac{1}{\epsilon} - \varphi' \right)(R_s + R_1) = - \frac{\kappa T \varphi' + 2 \varphi}{3 \varphi''}.
\end{equation}
Notice that the left hand side of this equation appears in (\ref{R_sterm}).  If we make the substitution on the right hand side of (\ref{newwaveeqn0}), we end up cancelling off all appearance of $R_s$, and eliminating the source term on the right hand side of our wave equation.  The final result is 
\begin{equation}\label{newwaveeqn}
\frac{\partial^2 R_1}{\partial t^2} - \frac{\partial^2 R_1}{\partial x^2} + \frac{1}{3 \varphi ''}\left( \frac{1}{\epsilon} - \varphi' \right) R_1 = 0,
\end{equation}
which is easy to solve numerically.

We can consider a wave moving from the left side of the transition point, $x < x_0$, toward the right.  I use $R_1 = e^{-(x - t)^2}$ for this initial wave, but this choice is arbitrary.  We want a form for $\rho$ near the boundary so that when $x > x_0$, $ \kappa \rho < R_0$.  We also want to make sure $R_1$ is much less than the change in $R$ over the domain we're considering.  We'll denote the domain boundaries by $(x_L, x_R)$, so we require $ \lvert \frac{R_1}{\kappa( \rho(x_L) - \rho(x_R))} \rvert \ll 1$.  Choosing $\rho$ to decrease by about $5\%$ from $x = 0$ to $x = x_0$ gives $\rho = -0.1 x/6 + 5.1$.

We these definitions, we're ready to choose a form for $f(R)$.  I'll use the analytic one in the motivation section.  To choose a good function for $\mathcal{F}(R)$ we have to be careful that the slope of $\mathcal{F}(R)$
at $R_0$ is less than $1$ in order to make sure the smoothed $f(R)$ has $f''(R) \geq 0$ as we approach $R_0$.  This ensures that the scalar mass remains positive, and so also makes sure that no instability arises as $R \rightarrow R_0$.  A form for $\mathcal{F}(R)$ that works is $$\mathcal{F}(R) = \frac{R_0}{1 + w}\left( 1 + w \left( \frac{R + \varepsilon}{R_0 + \varepsilon} \right)^2 \right),$$ where the term $\varepsilon = 0.04$ is used to suppress the negative second derivative of our function $g(R)$, and keep the total second derivative of $f(R)$ positive, $w = 100$ just helps make the second derivative of the quadratic term higher, to also help with this.  With these definitions, we can work out the effective mass term in equation (\ref{scalareffeqn}), $ \frac{1}{3 \varphi ''}\left( \frac{1}{\epsilon - \varphi'} \right)$.

We choose to reverse the effective theta function, $g(R)$, so that the $\varphi$ field is turned on at higher curvatures, and off at lower curvatures.  Then, our function $f(R)$ looks essentially the same as figure \ref{fig:5-3_1}.

\begin{figure}\label{fig:5-3_3}
\begin{center}
\includegraphics[width=0.5\textwidth]{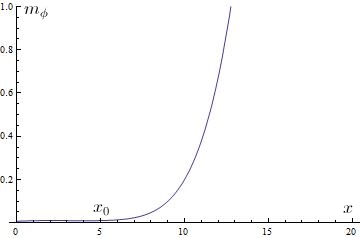}
\end{center}
\caption[Effective scalar field mass near critical points]{Effective mass for the scalar field.  It gets large past $x_0$, and the scalar can't propagate past this point.}  
\end{figure}

\begin{figure}\label{fig:5-3_3a}
\begin{center}
\includegraphics[width=0.5\textwidth]{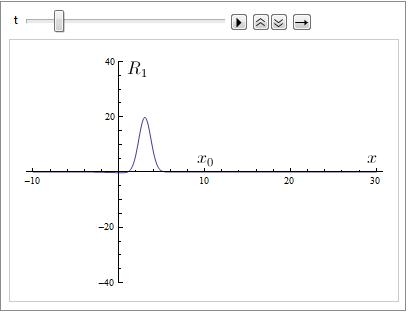}
\end{center}
\caption{A right-moving wave heads toward the boundary, near $x_0 = 10$.}  
\end{figure}

\begin{figure}\label{fig:5-3_3b}
\begin{center}
\includegraphics[width=0.5\textwidth]{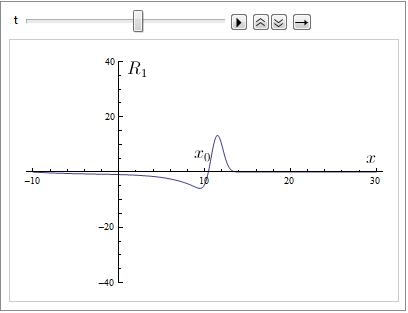}
\end{center}
\caption{The wave gets partially deflected as it approaches the boundary.}  
\end{figure}

\begin{figure}\label{fig:5-3_3c}
\begin{center}
\includegraphics[width=0.5\textwidth]{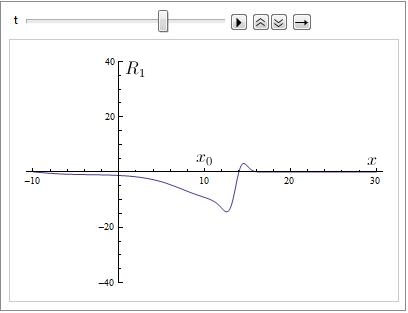}
\end{center}
\caption{The wave is deflected, and is now deformed and moving to the left.}  
\end{figure}

This results in an effective mass that increases toward smaller $R$ values, as depicted in figure \ref{fig:5-3_3}
, so as a scalar wave moves to the right, the mass begins increasing, and we get partial reflection as we continue moving right.  When we reach the boundary where the mass gets very large, the rest of the wave reflects.   This is illustrated in figures \ref{fig:5-3_3a} through \ref{fig:5-3_3c}.

Thus, we see that the waves can't pass from a high curvature region into a low curvature region.  The same argument applies in reverse if we reverse the theta function in $f(R)$, and cause the scalar field to be active in low curvature regions.  The result is, in the first case, if a wave is produced by a source in a high energy environment, then it cannot escape to a lower energy environment.  It is reflected by the boundary where $f''(R) = 0$.  

I should note that I have not been rigorous in effectively treating $f(R)$ with a Taylor expansion in the above analysis.  The function $f(R)$, since it uses the $arctan(x)$ function, inherits a finite radius of convergence within $ \lvert x \rvert < 1$.  Expanding around $R = R_0$, we find the radius of convergence to be $R < R_0(1 + \epsilon)$, or $-\epsilon R_0 < R < R_0(2 + \epsilon)$.  The density $\rho$ is such that $R_0$ stays well within these limits, but the perturbation I have chosen as an illustration is too large.  If I took into consideration how this perturbation changed the background $R_0$, then the Taylor expansion I have used would be invalid.

\section{Conclusions}
$\indent$Metric $f(R)$ gravity is a simple generalization of general relativity and a natural candidate for explaining dark energy and dark matter.  It is not motivated by first principles, but is instead a simple generalization of GR that is useful for exploring other gravitational effects.  It has appeared to violate solar system constraints, but that is simply because it has been investigated far from regions where $f''(R) \neq 0$.  It was thought that these regions are trivial, because the theory simply reduces to GR, but there are very interesting wave effects at the boundaries of these regions.  Further, the simplicity of reducing the theory to GR in these regions makes it a strong candidate for suppressing solar system effects of $f(R)$ gravity. 

I have shown that the effects of modifying gravity can, in principle, and at a classical level, be suppressed for certain ranges of scalar curvatures.  I've demonstrated a simple means for constructing toy $f(R)$ theories that turn on and off modifications to GR in certain ranges of $R$.  Before, the chameleon mechanism was used to suppress solar system effects, with unfortunate consequences for the equivalence principle.  One can now, in principle, devise an $f(R)$ that agrees with solar system constraints, and may still accounts for dark energy.  

To be more concrete, we can imagine a scenario where the transition scale $R_o$ is somewhere in between the background curvature of the interstellar medium and that of the intergalactic medium.  Then, we can have scalar gravitational waves propagating in the intergalactic medium, and also explain the accelerated expansion of the universe without use of a cosmological constant.  

We can imagine other interesting scenarios.  We can consider modifying gravity at larger curvatures try to account for dark matter.  We can have a situation where the transition scale is at a curvature (or background density) between that of the solar system and that of the interstellar medium.  Then, if we choose $f(R)$ to have a matter equation of state outside of the solar system, and to be like GR inside, we can potentially account for dark matter while agreeing with solar system constraints on gravity, all without having to make use of the chameleon mechanism.

A very interesting case would include two different transition scales.  We could have $f(R)$ suppressed for $R$ between solar system densities and interstellar medium densities, then take on a matter equation of state between interstellar densities and outer galactic densities, and finally take on a dark energy equation of state below galactic densities, in intergalactic space.  This would allow us to try to account for dark energy and dark matter while still agreeing with solar system constraints, and with non-trivial new wave effects at each of the boundaries.

\bibliographystyle{unsrt}
\bibliography{dissertation}

\begin{thebibliography}{10}

\bibitem{DeFelice2010}
Antonio De~Felice and Shinji Tsujikawa.
\newblock {f(R) theories}.
\newblock {\em Living Rev.Rel.}, 13:3, 2010.

\bibitem{Sotiriou}
Thomas Sotiriou and Valerio Faraoni.
\newblock {f(R) Theories of Gravity}.
\newblock {\em Rev. Mod. Phys.}, 82(1):451--497, 2010.

\bibitem{KhouryWeltman}
Justin Khoury and Amanda Weltman.
\newblock {Chameleon Cosmology}.
\newblock {\em Phys. Rev. D}, 69:044026, Feb 2004.

\bibitem{Lombriser2013}
Lucas Lombriser, Baojiu Li, Kazuya Koyama, and Gong-Bo Zhao.
\newblock {Modeling halo mass functions in chameleon f(R) gravity}.
\newblock {\em Phys.Rev.}, D87:123511, 2013.

\bibitem{Tsujikawa2008}
S.~Tsujikawa, K.~Uddin, S.~Mizuno, R.~Tavakol, and J.~Yokoyama.
\newblock Constraints on scalar-tensor models of dark energy from observational
  and local gravity tests.
\newblock {\em Phys. Rev. D}, 77, 2008.

\bibitem{Capozziello2008}
S.~Capozziello and S.~Tsujikawa.
\newblock Solar system and equivalence principle constraints on $f(r)$ gravity
  by chameleon approach.
\newblock {\em Phys. Rev. D}, 77, 2008.

\bibitem{HuiNicolisStubbs}
A.~Nicolis L.~Hui and C.~Stubbs.
\newblock {Equivalence Principle Implications of Modified Gravity Models}.
\newblock {\em Phys. Rev. D}, 80:104002, 2009.

\bibitem{Boehmer2007}
Christian~G. Boehmer, Tiberiu Harko, and Francisco~S.N. Lobo.
\newblock {Dark matter as a geometric effect in f(R) gravity}.
\newblock {\em Astropart.Phys.}, 29:386--392, 2008.

\bibitem{Lubini2011}
M.~Lubini, C.~Tortora, J.~Naf, Ph. Jetzer, and S.~Capozziello.
\newblock {Probing the dark matter issue in f(R)-gravity via gravitational
  lensing}.
\newblock {\em Eur.Phys.J.}, C71:1834, 2011.

\bibitem{Capoziello2012}
S.~Capozziello and M.~De~Laurentis.
\newblock The dark matter problem from f(r) gravity viewpoint.
\newblock {\em Annalen der Physik}, 524(9-10):545--578, 2012.

\bibitem{Mendoza2006}
Sergio Mendoza and Y.M. Rosas-Guevara.
\newblock {Gravitational waves and lensing of the metric theory proposed by
  Sobouti}.
\newblock {\em Astron.Astrophys.}, 472:367--371, 2007.

\bibitem{Faraoni2006}
Valerio Faraoni.
\newblock {Matter Instability in Modified Gravity}.
\newblock {\em Phys.Rev.}, D74:104017, 2006.

\end{thebibliography}

\end{document}